\documentclass[a4paper,11pt]{article}
\pdfoutput=1 

\usepackage{jcappub} 
\usepackage{graphicx}
\usepackage{subcaption}
\usepackage{float}

\usepackage[T1]{fontenc} 

\title{\boldmath Constraining neutrino mass with weak lensing Minkowski Functionals}
\author[a,1]{Gabriela A. Marques,\note{Corresponding author.}}
\author[b]{Jia Liu,}
\author[c]{Jos\'e Manuel Zorrilla Matilla,} 
\author[c]{Zolt\'an Haiman,}
\author[a]{Armando Bernui}
\author[a]{and Camila P. Novaes}

\affiliation[a]{Observat\'orio Nacional, Rua General Jos\'e Cristino 77, S\~ao Crist\'ov\~ao, 20921-400 Rio de Janeiro, RJ, Brazil}
\affiliation[b]{Department of Astrophysical Sciences, Princeton University, Princeton, NJ 08544, USA}
\affiliation[c]{Department of Astronomy, Columbia University, New York, NY 10027, USA}
\emailAdd{gabrielamarques@on.br}

\abstract{The presence of massive neutrinos affects structure formation, leaving imprints on large-scale structure observables such as the weak lensing field. The common lensing analyses with two-point statistics are insensitive to the large amount of non-Gaussian information in the density field. We investigate non-Gaussian tools,  in particular the Minkowski Functionals (MFs)---morphological descriptors including area, perimeter, and genus---in an attempt to recover the higher-order information. We use convergence maps from the Cosmological Massive Neutrino Simulations (\texttt{MassiveNus}) and assume galaxy noise, density, and redshift distribution for an LSST-like survey. We show that MFs are sensitive to the neutrino mass sum, and the sensitivity is redshift dependent and is non-Gaussian. We find that redshift tomography significantly improves the constraints on neutrino mass for MFs, compared to the improvements for the power spectrum.  We attribute this to the stronger redshift dependence of neutrino effects on small scales. We then build an emulator to model the power spectrum and MFs, and study the constraints on $[M_{\nu}$, $\Omega_{m}$, $A_{s}]$ from the power spectrum, MFs, and their combination. We show that MFs significantly outperform the power spectrum in constraining neutrino mass, by more than a factor of four. However, a thorough study of the impact from systematics such as baryon physics and galaxy shape and redshift biases will be important to realize the full potential of MFs.}

\begin{document}
\maketitle
\flushbottom
\section{Introduction}
\label{sec:intro}

The discovery of neutrino oscillation implies that neutrinos have non-zero masses~\cite{becker1992electron, fukuda1998measurements,ahmed2004measurement}. From particle experiments measuring the differences of neutrino mass squared, we can obtain a lower bound of the neutrino mass sum $M_{\nu}>$0.06~eV (0.1~eV) assuming a normal (inverted) hierarchy. However, the upper bound remains unconstrained by these oscillation experiments. 
The expansion history and structure formation are modified by the presence of massive cosmic neutrinos---relic neutrinos decoupled from the rest of matter when the temperature of the universe was a few MeV. With large thermal pressure, cosmic neutrinos suppress the matter clustering on scales smaller than their free-streaming length. Since the level of suppression depends on the neutrino masses, large-scale structure was hence proposed to be a powerful tool to constrain $M_{\nu}$~\cite{seljak2005cosmological,lesgourgues2006massive}.    

At present, upper limits  on $M_{\nu}$ from cosmological observations are already approaching the lower limit. For example, the most recent data of Cosmic Microwave Background (CMB) temperature and polarization combined with baryon acoustic oscillations (BAO) and CMB lensing data put an upper limit on $M_{\nu}$ $< 0.12$~eV (at 95$\%$ C.L.) \cite{aghanim2018planck}. This number will soon be significantly improved by upcoming galaxy surveys such as the LSST\footnote{Large Synoptic Survey Telescope: \url{http://www.lsst.org}}~\cite{abellpa}, WFIRST\footnote{Wide-Field Infrared Survey Telescope: \url{http://wfirst.gsfc.nasa.gov} }, and Euclid\footnote{Euclid: \url{http://sci.esa.int/euclid}} and CMB surveys such as the Simons Observatory\footnote{Simons Observatory: \url{https://simonsobservatory.org}}~\cite{ade2018simons} and CMB-S4\footnote{CMB-S4: \url{https://cmb-s4.org/}}~\cite{abazajian2016cmb}.
 
Significant improvements on neutrino mass constraints are expected to come from weak lensing (WL), where background galaxies are distorted by foreground matter and are hence used to map out the total matter distribution in our universe~\cite{bartelmann2001weak,hoekstra2008weak}. In the past decade, several pioneering WL surveys went online and achieved statistically significant constraints on cosmology~\cite{heymans2012cfhtlens, kids2016,mandelbaum2017weak,abbott2018dark}.

So far, most of the WL analyses are done using second-order statistics, the two-point correlation function (real space) or the power spectrum (Fourier space). These two-point statistics can capture all the information only if the field is Gaussian. However, matter clustering is highly nonlinear, and hence non-Gaussian, on scales $\lesssim$10~Mpc. Massive neutrino signals are also the largest on these small scales. Thus, non-Gaussian statistical tools are necessary in order to capture the missing information and to constraint $M_{\nu}$. 
 
In this work, we forecast the constraints on $M_{\nu}$ from non-Gaussian statistics, in particular from the Minkowski Functionals~(MFs). MFs are a set of morphological descriptors, first introduced to cosmology by~Ref.~\cite{mecke288k} and since then used as a tool to detect deviations from Gaussianity~ \cite{hikage2006primordial,park2005topology,komatsu2009five, ducout2012non, novaes2014searching, novaes2015neural, lewis2016planck}. Studies of MFs with WL simulations~\cite{munshi2016lensing,fang2017new} and data~\cite{petri2015petri,shirasaki2014statistical} found that MFs are potentially more sensitive than two-point statistics in constraining $\Omega_{m}$ (matter density) and $\sigma_{8}$ (r.m.s. linear matter fluctuation on the scale of 8~Mpc/$h$), specially when multiple redshifts and smoothing scales are combined~\cite{kratochvil2012probing}.

Motivated by previous findings, we perform a detailed study of WL MFs and power spectrum for an LSST-like survey, with a focus on constraining $M_\nu$. We use simulated convergence maps from the Cosmological Massive Neutrino Simulations 
(\texttt{MassiveNuS})\footnote{The \texttt{MassiveNuS} data products, including snapshots, halo catalogues, merger trees, and galaxy and CMB lensing convergence maps, are publicly available at \url{http://ColumbiaLensing.org}.} 
which include 101 models with three varying cosmological parameters, $M_{\nu}$, $\Omega_{m}$ and $A_{s}$ (primordial power spectrum amplitude) and five tomographic source planes at $z_s$=0.5--2.5. We build an emulator to predict MFs at an arbitrary cosmology and use Markov chain Monte Carlo to sample the likelihood. We investigate the constraints from each of the MFs, from adding evolution information (i.e. tomography vs single-redshift), and from the combination of the power spectrum and MFs.

This paper is organized as follows. First, we review the power spectrum and MFs in the context of WL convergence in section~\ref{sec:theory}. Next, we describe the \texttt{MassiveNuS} suite in section~\ref{sec:simulations} and our analyses in section~\ref{sec:analysis}. We present our findings and discuss their implications in section ~\ref{sec:results}. Finally, we conclude in section~\ref{sec:conclusions}.

\section{Background}
\label{sec:theory}

\subsection{Weak lensing convergence}
Matter  along the line-of-sight deflects the path of photons emitted by background galaxies, giving rise to the phenomenon of gravitational lensing. In the weak regime, lensing slightly distorts (shears) and (de-)magnifies the source image. The lensing convergence $\kappa$, describing the amplitude of such magnification, can be expressed as a weighted projection of the mass overdensity using Born approximation,

\begin{eqnarray}
\kappa(\mathbf{\hat{n}}) &=& \int_{0}^{\infty} dz W(z)\delta(\chi(z)\mathbf{\hat{n}},z),
\end{eqnarray}
where $\delta$ is the projected matter overdensity, $\chi(z)$ is the comoving distance to redshift $z$, and $W$ is the lensing kernel defined as 
\begin{eqnarray}
W(z) &=& \frac{3\Omega_{m}}{2c}\frac{H_{0}^2}{H(z)}(1+z)\chi(z)       \int_{z}^{\infty} \frac{dn(z_{s})}{dz_{s}}\frac{\chi(z_{s})-\chi(z)}{\chi(z_{s})} dz_{s},
\label{eq:kappa}
\end{eqnarray} 
where $z_{s}$ is the source redshift, $dn/dz_{s}$ is the source redshift distribution, and $H(z)$ is the Hubble parameter with  present-day value $H_{0}$. 
 
\subsection{Power spectrum}
WL studies commonly use two-point statistics to constrain cosmology. In this work, we use the Fourier transform of the two-point correlation function, the power spectrum. The convergence power spectrum $C_{l}^{\kappa}$ is defined as
\begin{eqnarray}
\langle \hat{\kappa}(\mathbf{l})\hat{\kappa}(\mathbf{l'})\rangle = (2\pi)^2\delta_{D}(\mathbf{l+l'})C_{\ell}^{\kappa},
\end{eqnarray}
where $\delta_{D}$ is the Dirac delta function and $\hat{\kappa}(\mathbf{l})$ is the Fourier transform of the two-dimensional~(2D) $\kappa$ field. Under the Limber approximation, the power spectrum can be expressed as a weighted line-of-sight integral over the three-dimensional matter power spectrum $P(k,z)$   
\begin{eqnarray}
C_{l}^{\kappa} &=& \int_{0}^{z_{s}} \frac{dz}{c}\frac{H(z)}{\chi^2(z)}W^2(z)P\bigg(k=\frac{\ell}{\chi(z)},z \bigg).
\label{eq:clkappa}
\end{eqnarray}
 
\subsection{Minkowski Functionals}
\label{sec:MFs}
Minkowski Functionals are a set of morphological descriptors  invariant under rotations and translations~\cite{hadwiger1957,hikage2008effect,mecke288k,ducout2012non}. The morphological properties of a given random field in a $D$-dimensional space can be completely characterized by a set of $D+1$ functionals. For a 2D convergence field $\kappa(\mathbf{\hat{n}}$), there are three MFs---the area $V_0$, perimeter $V_1$, and genus $V_2$,
\begin{eqnarray}
V_0\left(\nu\right) &=& \frac{1}{A} \int_{\Sigma\left(\nu\right)}  da , \\
V_1\left(\nu\right) &=& \frac{1}{4A} \int_{\partial \Sigma \left(\nu\right)} dl , \\
V_2\left(\nu\right) &=& \frac{1}{2\pi A} \int_{\partial \Sigma\left(\nu\right)} \mathcal{K} dl,
\end{eqnarray}
where $da$ and $dl$ are the area and length elements. The MFs are defined for an excursion set $\Sigma \left( \nu \right)=\{ \kappa > \nu \sigma_0 \}$, that is, the set of pixels with $\kappa$ values exceeding the threshold level $\nu\sigma_0$, where $\sigma_0$ is the standard deviation of the field. The boundary of the excursion set is $\partial \Sigma\left(\nu\right) = \{\kappa = \nu \sigma_0\}$. The first MF $V_0$ describes the area covered by the excursion set, equivalent to the cumulative distribution function. The second MF $V_1$ describes the length of the boundary. The third MF $V_2$ describes  the total number of connected regions (``islands'') above the threshold minus that below (``holes''), or the integrated geodesic curvature $\mathcal{K}$ along the boundary. 

The MFs capture statistical information at all orders of correlation, making them useful probes of non-Gaussianity. They can be evaluated analytically for Gaussian random fields (GRFs) and weakly non-Gaussian fields~\cite{tomita1986curvature,matsubara2010analytic}. The accuracy of the analytical predictions have been validated against simulations~\cite{kratochvil2012probing}. However, for highly nonlinear fields, such as those expected from upcoming WL surveys at small angular scales, analytical predictions break down~\cite{petri2013cosmology}. Therefore, we use numerical simulations to model MFs.

\section{Simulation}
\label{sec:simulations}
In this work, we use convergence maps from the Cosmological Massive Neutrino Simulations (\texttt{MassiveNus}). Here we describe briefly the set-up of the N-body ray-tracing simulations. More details on code validations can be found in the simulation paper~\cite{liu2018massivenus}.

\subsection{MassiveNuS: N-body ray-tracing simulations}

\texttt{MassiveNus} consists of 101 models in flat-$\Lambda$CDM cosmology, with three varying parameters: total mass of neutrinos $M_{\nu}$, matter density $\Omega_{m}$, and primordial power spectrum amplitude $A_{s}$. These parameters are sampled randomly following the Latin Hyper Cube algorithm in the range of $M_{\nu} = [0.0, 0.62]$~eV, $\Omega_{m}= [ 0.18, 0.42]$, and $A_{s}\times 10^{9} = [1.29, 2.91]$.  
The N-body simulations use the public tree-Particle Mesh code \texttt{Gadget-2}\footnote{\url{http://www.mpa-garching.mpg.de/gadget}} \cite{springel2005cosmological}, with a box size of $512$ Mpc/$h$ and $1024^{3}$ cold dark matter particles. The background density of neutrinos is evolved using a fast linear response algorithm \texttt{kspace neutrinos}~\cite{ali2012efficient, bird2018}. The simulations accurately capture  the impact of massive neutrinos on structure growth at sub-percent precision for $k< 10$ $h$/Mpc. 

The convergence maps are generated for five galaxy source planes $z_{s}= 0.5, 1.0, 1.5, 2.0, 2.5$, using the public ray-tracing code \texttt{LensTools} \cite{petri2016}. We do not assume small angle or undisturbed geodesics, as Ref.~\cite{petriBornapprox} found that, despite sufficient for the power spectrum, the Born approximation can bias non-Gaussian statistics.
For each cosmological model and redshift bin, 10,000 realizations are generated by randomly rotating and shifting the lens planes. In order to avoid correlations between the noise in our model and  covariance ~\cite{carron2013assumption}, we also generate an additional set of 10,000 maps with different initial conditions for the fiducial, massless model ($M_{\nu}$ =0.0, $\Omega_{m}$ = 0.3, $A_{s} = 2.1\times 10^{-9}$). Each map has $512^2$ pixels and is $12.2$ deg$^2$ in size, and hence has a spatial resolution $\approx 0.4$ arcmin per pixel.

\subsection{LSST-like mocks}

Using these convergence maps, we next generate \texttt{LSST}-like mocks. We follow the configurations in the \texttt{LSST} Science book~\cite{abellpa}, with galaxy number density $n_{gal} = 50$ arcmin$^{-2}$ and source redshift distribution,
\begin{eqnarray}
n(z) \varpropto z^{2} \exp(-2z).
\end{eqnarray}
Assuming a redshift bin width $\Delta z_s=0.5$, for each source redshift at $z_s = 0.5, 1.0, 1.5, 2.0, 2.5$, we obtain a number density $n_{gal} = 8.83, 13.25, 11.15, 7.36, 4.26$ arcmin$^{-2}$, respectively.  We discard galaxies at $z_{s} < 0.25$ and $z_{s} > 2.75$, resulting in a smaller total number density of $44.85$ arcmin$^{-2}$. To assess the utility of tomography, i.e. joint analysis of multiple redshift bins, we also consider a set of maps with a single source plane $z_{s}= 1.0$, but with the galaxy density equivalent to the sum of all galaxies in the five tomographic bins.
 
We then add Gaussian noise to each pixel with variance, 

\begin{eqnarray}
\sigma_{noise}^2= \frac{\langle \sigma_{\lambda}^{2} \rangle }{n_{gal}\Delta\Omega}
\end{eqnarray}
where $\sigma_{\lambda}=0.35$ is the shape noise and $\Delta\Omega=0.17$ arcmin$^2$ is the solid angle of a pixel. For each realization, the maps at five redshifts are correlated in signal, as the light rays go through the same large-scale structure, but uncorrelated in noise.

\section{Analysis}
\label{sec:analysis}

\subsection{Statistical measurements}

The convergence power spectrum and three MFs are measured for each of the 101 (models) $\times$ 5 (source redshifts) $\times$ 10,000 (realizations) convergence maps. The power spectrum $C_{\ell}^{\kappa}$, is evaluated at 25 logarithmically-spaced bins in the range $300 <\ell< 5000$.

Before measuring the MFs, each map is smoothed with a one arcmin Gaussian filter. 
All three MFs are measured over 50 uniformly-spaced bins, between $\nu=-6.0$ and $\nu=6.0$ in units of the standard deviation of the shape noise, where $\sigma_{noise}$ =[0.034, 0.027, 0.030, 0.037, 0.048] for $z_{s}$=[0.5, 1.0, 1.5, 2.0, 2.5], respectively. For the single redshift case, the galaxy shape noise is $\sigma_{noise}= 0.0157$. 

We found that bins at MF tails, while may contain additional information, are too noisy to be modeled accurately within our current computational capability. Therefore, in the likelihood analysis, we discard bins at both low and high tails of the distribution. To be specific, we remove the extreme bins with absolute value less than 5$\%$ of the maximum value.
For $V_0$, we convert the distribution to a probability function before applying this criterion.

\subsection{Emulator based on Gaussian Processes}

To model the statistics at an arbitrary point in parameter space, we built an emulator based on measurements from the noise included mocks in 101 models. We interpolate in the 3D parameter space using a Gaussian Process (GP) with a squared exponential kernel. We use the GP algorithm implemented in the \texttt{scikit-learn}\footnote{\url{http://scikit-learn.org}} Python package. We test the accuracy of our GP emulator by comparing the prediction with the true value, where the prediction is made by a test emulator built using all but the model in interest. We find the deviations from the true values consistently at sub-percent level or lower and significantly smaller than the statistical error expected from the shape noise. More detailed tests of the emulator can be found in our companion papers investigating other non-Gaussian statistics~\cite{coulton2018,li2018constraining}.

\subsection{Full covariance}

We use an independent set of $N_r$=10,000 simulations at the fiducial massless model to evaluate the covariance matrix,
\begin{eqnarray}
\mathrm{C}_{ij} &=& \frac{1}{N_r-1}\sum_{n=1}^{N_r}(d_{i}^{n}- \langle d_{i}\rangle)(d_{j}^{n}- \langle d_{j}\rangle),
\label{eq:cov}
\end{eqnarray}
where $d_{i}^{n}$ denotes the $i^{th}$ bin and $n^{th}$ realization for the statistical descriptor $d$, which can be the power spectrum, one of the MFs, or a combination of them in the case of joint analysis. We use a cosmology-independent covariance, which is necessary to obtain unbiased constraints when assuming a Gaussian likelihood (see section~\ref{sec:likelihood}). Compared with our 12.25 deg$^2$ simulated maps, LSST will cover roughly 20,000 deg$^2$ sky~\cite{abellpa}. To roughly account for this difference in sky coverage, we scale the simulated covariance by a ratio=$12.25/20,000$.\

Fig.~\ref{fig:covs} shows the correlation coefficients of the full covariance matrix of the three MFs and power spectrum (``PS'') in  5 redshifts bins for both the noiseless (left) and noisy (right) maps,
\begin{eqnarray}
\rho_{ij} &=& \frac{\mathrm{C}_{ij}}{\sqrt{\mathrm{C}_{ii}\mathrm{C}_{jj}}},
\end{eqnarray}
where $i$ and $j$ indicate the bin number. The total numbers of bins are different for the noisy and noiseless maps for the same selection criterion. In the noiseless (noisy) case, we have 70, 53, 53, and 125 (70, 120, 130, and 125) for $V_{0}$, $V_{1}$, $V_{2}$, and power spectrum, respectively. Within each statistical block, 5 sub-blocks represent the five redshift bins.\ 

\begin{figure*}
    \centering
    \begin{subfigure}[b]{0.5\linewidth}
        \centering
        \includegraphics[scale= 0.2]{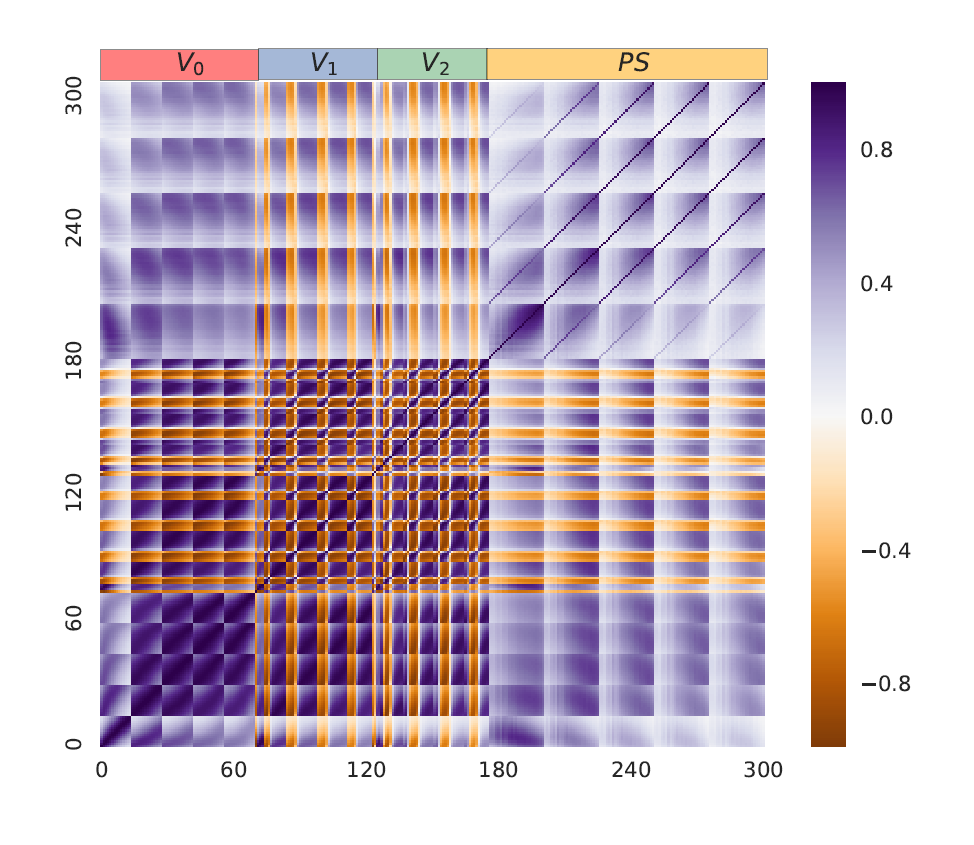}
    \end{subfigure}%
    ~ 
    \begin{subfigure}[b]{0.5\linewidth}
        \centering
        \includegraphics[scale= 0.2]{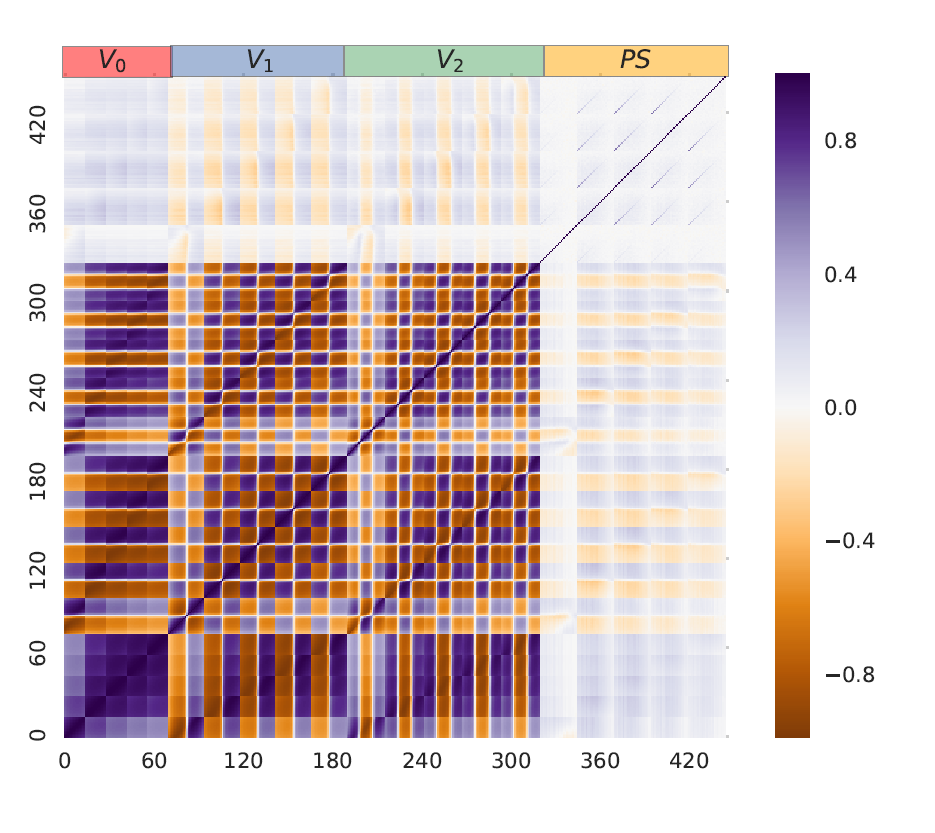}
        \label{fig:covs}
    \end{subfigure}
\caption{Noiseless ({\bf left}) and noisy ({\bf right}) correlation coefficients of the full covariance matrix of the three Minkowski Functionals ($V_{0}$, $V_{1}$, $V_{2}$) and the power spectrum (PS). Each of the four statistic blocks contains five redshift sub-blocks. The cross-covariance (or the off-diagonal terms), showing non-trivial correlation, are included in our likelihood analysis. }
    \label{fig:covs}
\end{figure*}

For the noiseless power spectrum, contribution of the off-diagonal terms are larger at lower redshifts and high $\ell$ bins, and it is significantly reduced in the noisy case. However, MFs in both the noiseless and noisy cases show non-trivial off-diagonal components, indicating the importance to model the full covariance for the MFs or the joint analysis.

\subsection{Parameter Estimation}
\label{sec:likelihood}

To forecast constraints on the cosmological parameters $\boldsymbol{\theta}=[\Omega_m, A_s, M_{\nu}]$ from a given statistical estimator (the three MFs, power spectrum, or a combination of them), we assume a Gaussian likelihood,

\begin{equation}
P\left(\boldsymbol{\theta}|\mathbf{x}\right) \varpropto
\exp \left[-\frac{1}{2} \left(\mathrm{x}_i-\mathrm{\mu}_i\left(\boldsymbol{\theta}\right)\right) \mathrm{C}_{ij}^{-1} \left(\mathrm{x}_j-\mathrm{\mu}_j\left(\boldsymbol{\theta}\right)\right)\right]
\label{eq:likelihood}
\end{equation}
where $\mathbf{x}$ is the ``observed data'', which we assume the average of the statistic(s) of interest at the fiducial massive model, and $\boldsymbol{\mu}(\boldsymbol{\theta})$ is the model prediction  computed using our emulator.
Although the estimated covariance from simulations (eq.~\ref{eq:cov}) is unbiased, its inverse can be biased due to the limited number of realizations. We correct the inverse of covariance following Ref.~\cite{hartlap2007your},
\begin{equation}
\mathrm{C}^{-1} = \frac{N_r-N_{b}-2}{N-1}\mathrm{C}^{-1}_{*},
\end{equation}
where $N_{b}$ is the number of bins, $N_r$=10,000 is the number of realizations and $C_{*}$ is the covariance computed from the simulations. We adopt a cosmology-independent covariance,  which is necessary when one assumes a Gaussian likelihood (as done in this work). Ref.~\cite{Carron2013} showed that by assuming a Gaussian likelihood, one breaks the correlation between the model (or the mean) and covariance and hence assigns the information content to the model. Under this framework, using a cosmology-dependent covariance would double count the cosmological information, and hence artificially shrink the error. 

We sample the posterior likelihood using Markov Chain Monte Carlo
~\cite{foreman2013emcee}. We use a flat prior and assume a nonzero $M_{\nu}$. Our results are stable against the length of the chain as well as the initial walker positions.

\section{Results and Discussions}
\label{sec:results}

\subsection{Neutrino signature in the MFs}

\begin{figure}[!h]
\centering
\begin{subfigure}{1.5\textwidth}
   \includegraphics[scale= 0.65]{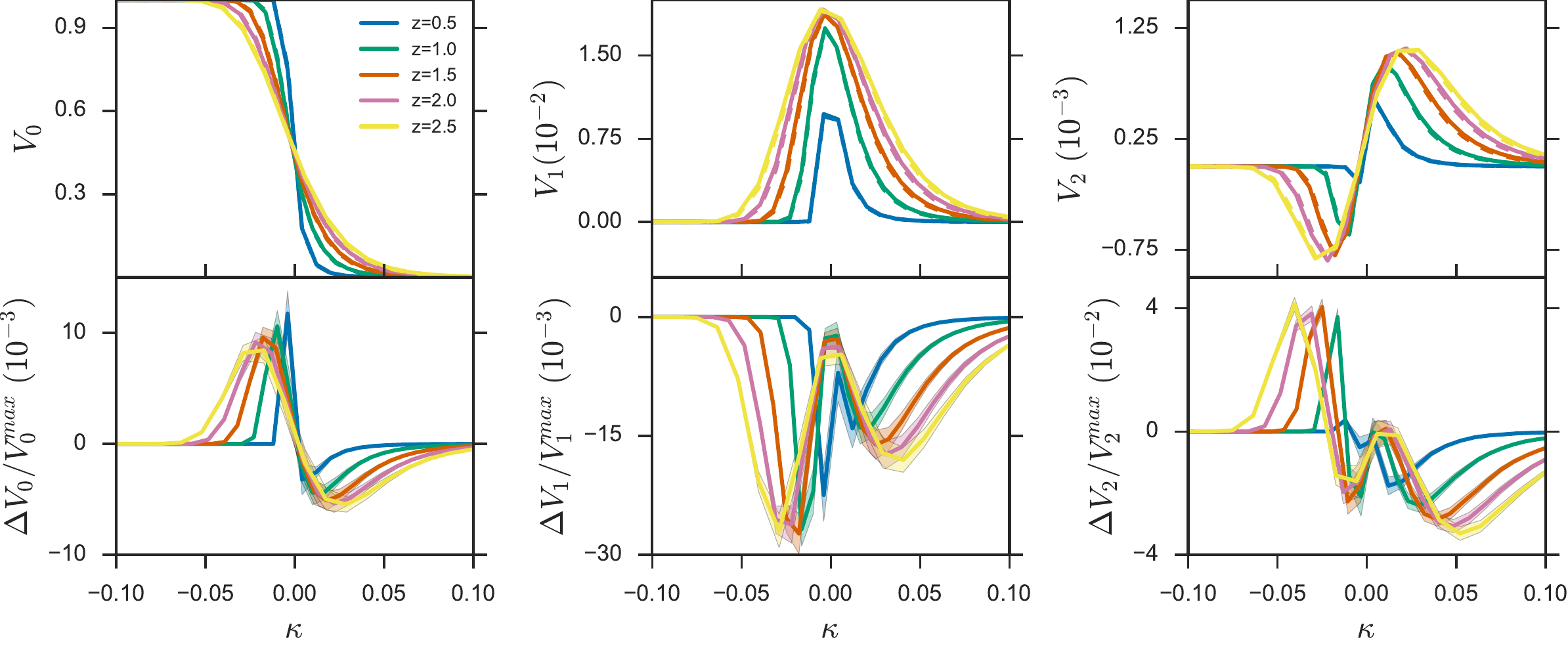}
\end{subfigure}
\centering
\begin{subfigure}{1.8\textwidth}
   \includegraphics[scale= 0.65]{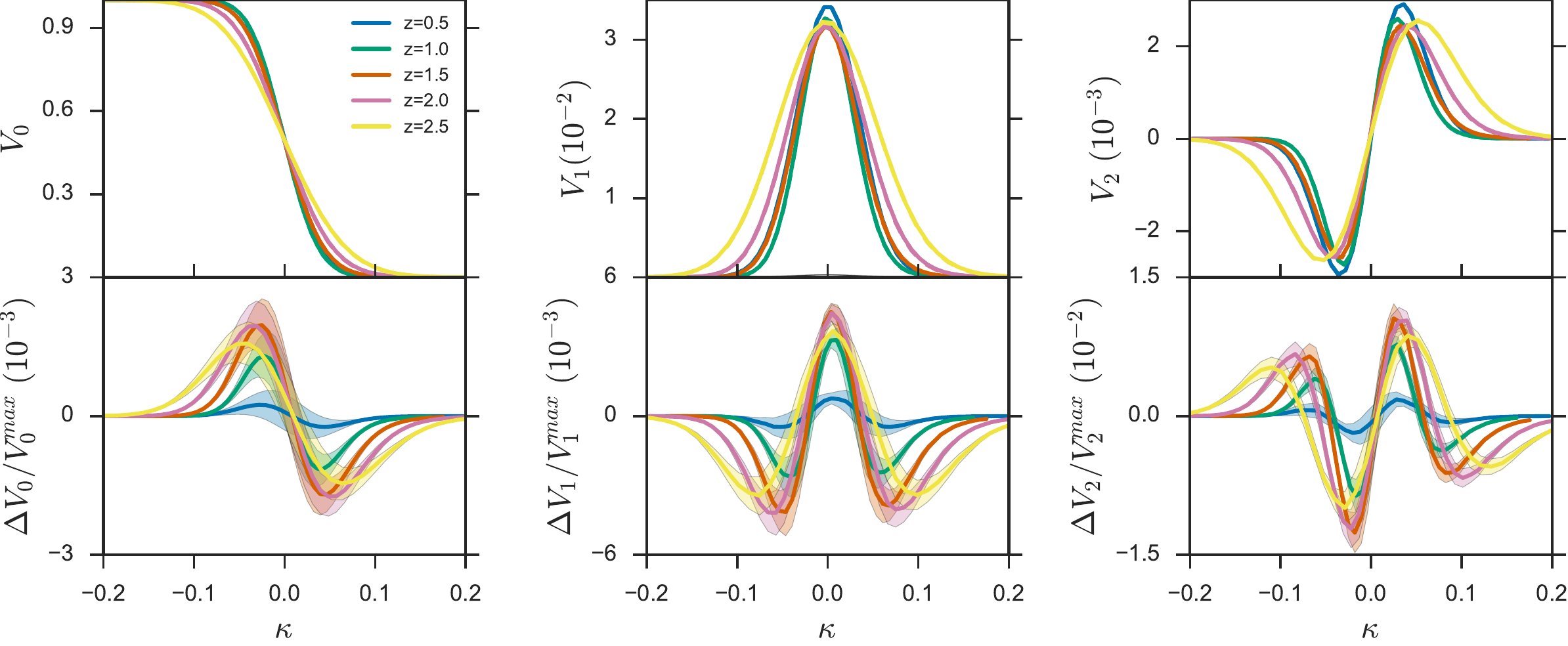}
\end{subfigure}
\caption{Three Minkowski Functionals, $V_{0}$ ({\bf left}), $V_{1}$ ({\bf middle}) and $V_{2}$ ({\bf right}) as a function of the threshold $\kappa$ for the massive (dashed) and massless (solid) fiducial models (top panels), as well as their fractional differences (bottom panels), at five source redshifts $z_s$=[0.5, 1.0, 1.5, 2.0, 2.5] for the noiseless ({\bf top row}) and noisy ({\bf bottom row}) cases. The expected LSST errors are shown as shaded color bands.}
\label{fig:MFs_signature}
\end{figure}

\begin{figure}[!h]
\includegraphics[width=1.05\textwidth]{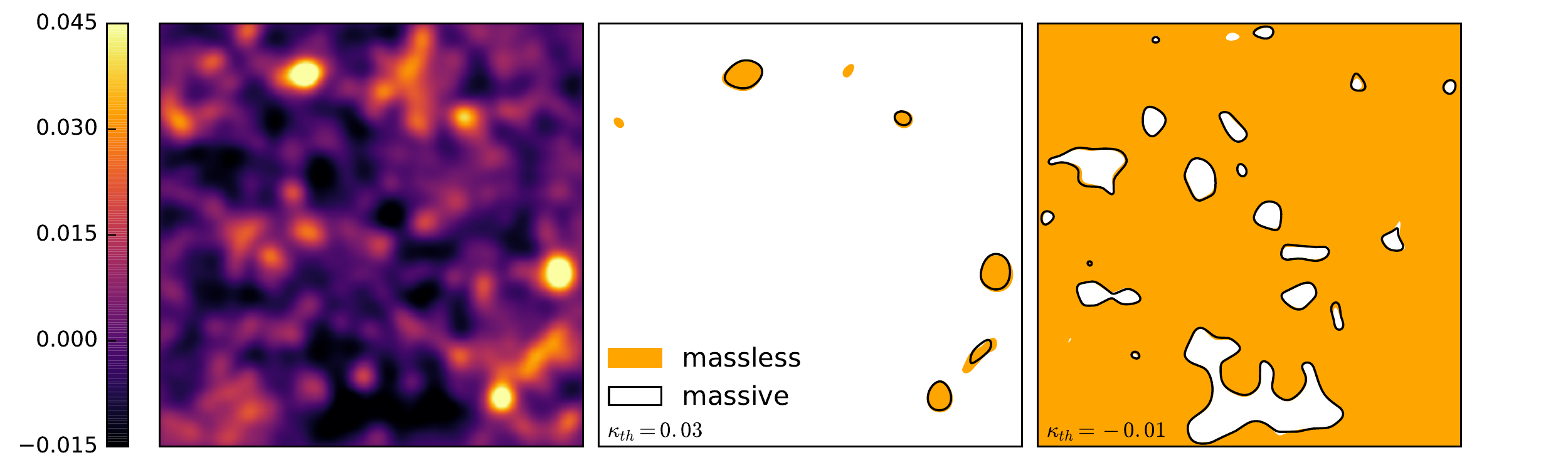}
\caption{{\bf Left}: a $z_s=1$ sample convergence map (0.77 deg$^2$) at the  fiducial massless model, smoothed with a one arcmin Gaussian window. {\bf Middle:} the excursion set (shaded orange regions) at a high threshold $\kappa_{th}=0.03$. We also show that for the fiducial massive (0.1~eV)  model (black lines, without shades for clarity), where other parameters are held fixed. {\bf Right:} Similar to the middle panel, but for a low threshold $\kappa_{th}=-0.01$.}
\label{fig:sample_map}
\end{figure}

To understand the effect of massive neutrinos,  we compare the average MFs in the fiducial massive model with that in the fiducial massless model, where $M_{\nu}= 0.1$~eV and $M_{\nu}= 0.0$~eV, respectively, with all other cosmological parameters fixed.  We show in Fig.~\ref{fig:MFs_signature} the average $V_{0}$, $V_{1}$ and $V_{2}$ for both models at five source redshifts, as well as their differences normalized by the maximum value. 
We show both noiseless (upper row) and noisy (lower row) cases. We also include the expected LSST (68\% C.L.) errors in shaded bands. To visualize the effect of massive neutrinos, we show in Fig.~\ref{fig:sample_map} a sample map threshed at two different levels $\kappa_{th}$=0.03 and $\kappa_{th}$=$-0.01$, for both fiducial models.

Focusing on the noiseless case, where the physical effects are more transparent, we find that massive neutrinos suppress the high $\kappa$ tails in $V_0$, thought to be associated with the most massive halos in the universe. This matches the expectation that massive neutrinos suppress the growth of large-scale structure. 
Similarly, at high $\kappa_{th}$, fewer regions in massive neutrino cosmology can pass the threshold, resulting in smaller perimeters ($V_1$) around these regions and a smaller number of ``islands'' (i.e. reduced $V_2$). We observe more complicated behavior at low $\kappa_{th}$ (Fig.~\ref{fig:sample_map} right panel), due to the non-spherical nature of under-dense regions. When galaxy noise is added (Fig.~\ref{fig:MFs_signature} lower panels), these features remain, albeit with lower amplitude.

\subsection{Comparison to Gaussian predictions} 
\begin{figure}[!htbp]
	\centering
		\begin{subfigure}{0.49\linewidth}
	\includegraphics[width=\linewidth]{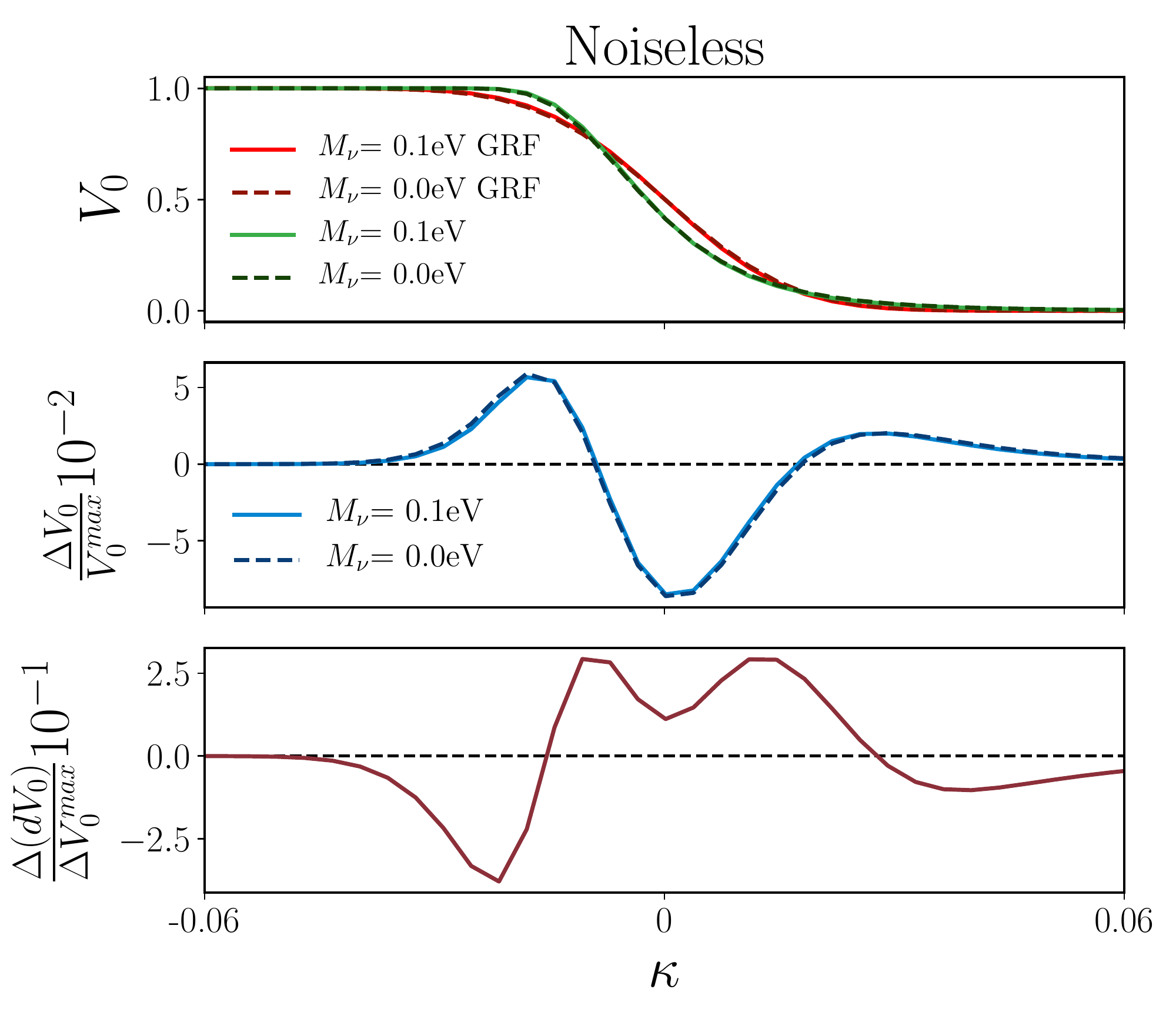}
	\end{subfigure}	
	\begin{subfigure}{0.49\linewidth}
		\includegraphics[width=\linewidth]{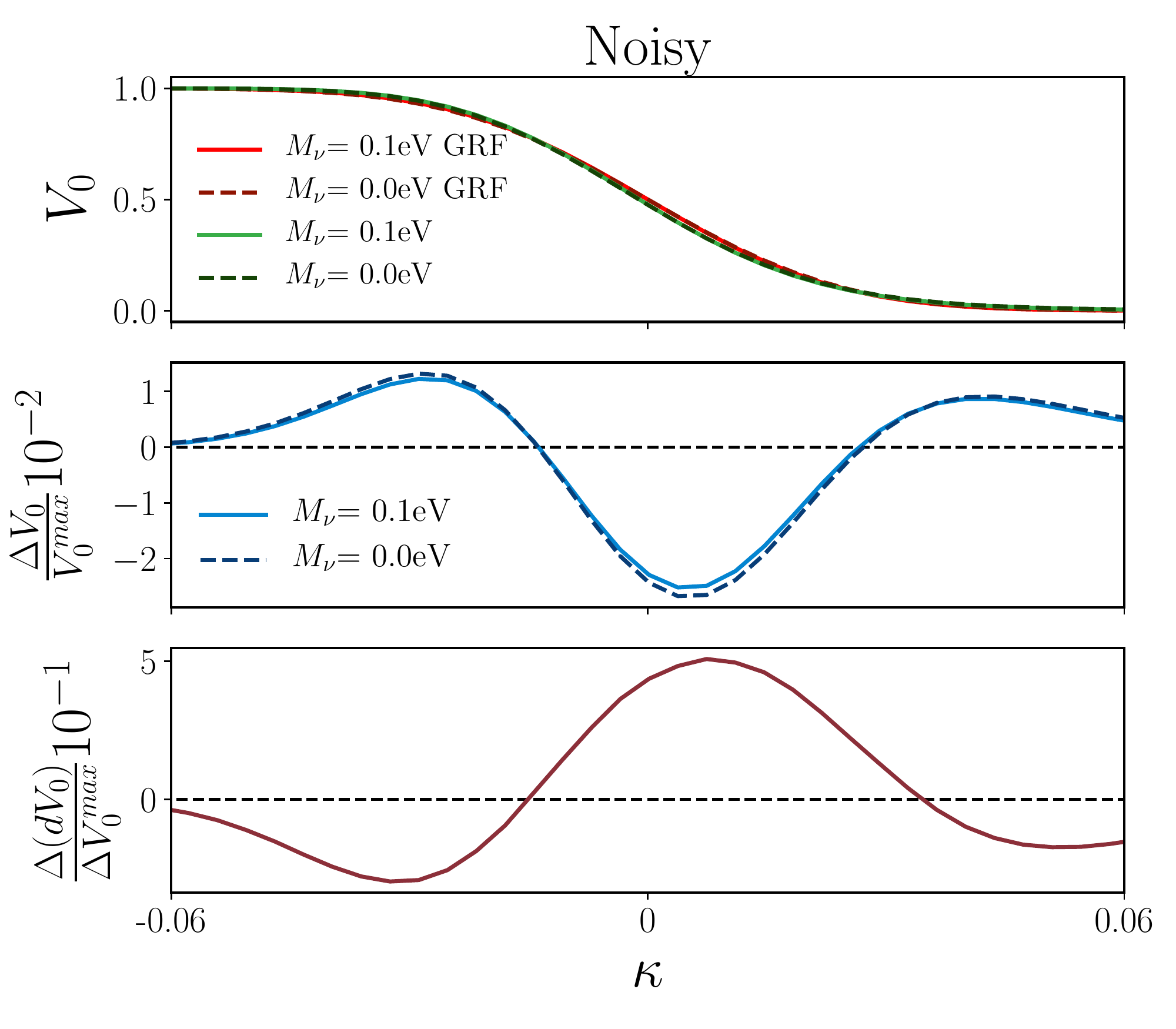}
	\end{subfigure}	
	\begin{subfigure}{0.49\linewidth}
		\includegraphics[width=\linewidth]{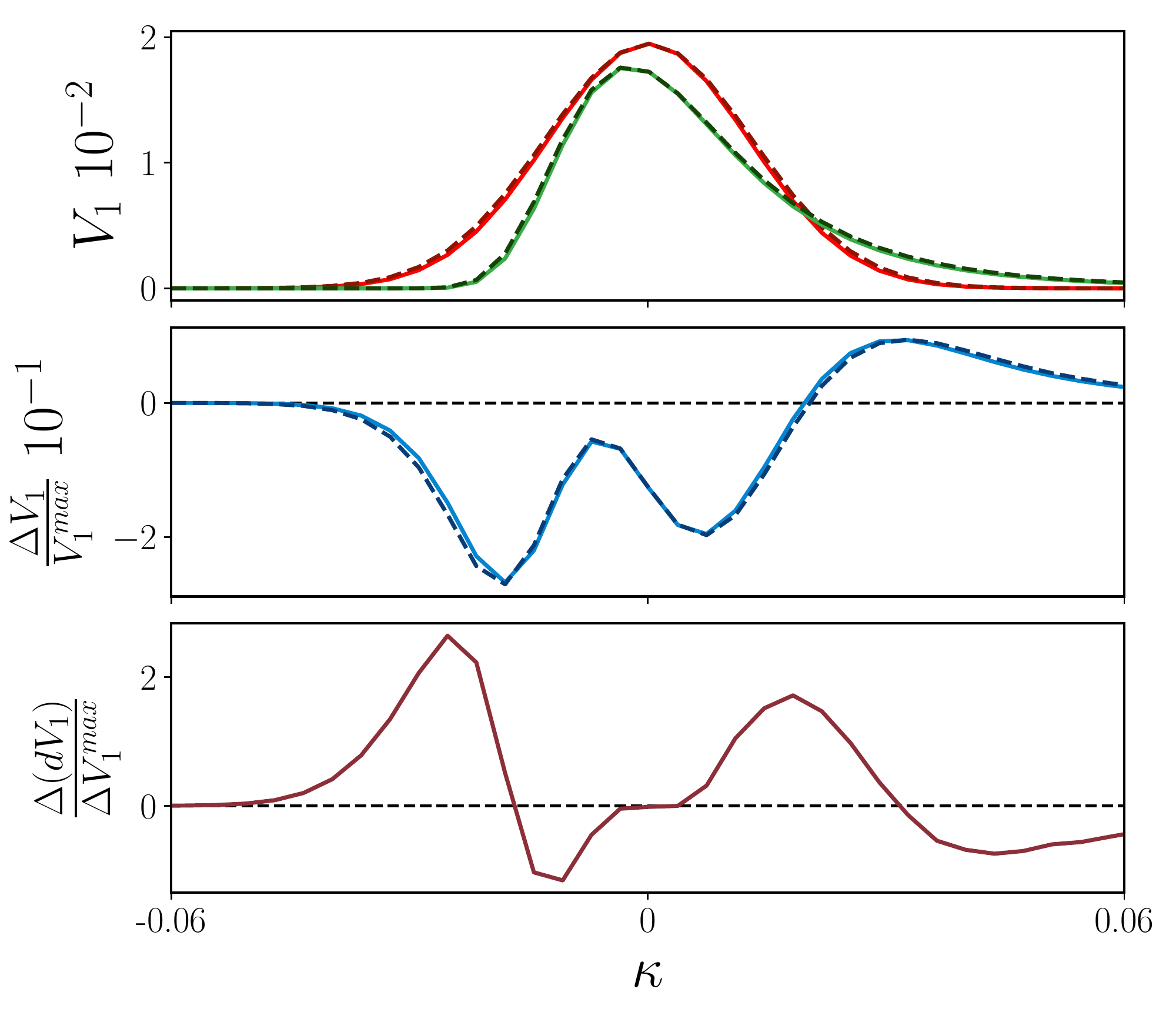}
	\end{subfigure}	
	\begin{subfigure}{0.49\linewidth}
	\includegraphics[width=\linewidth]{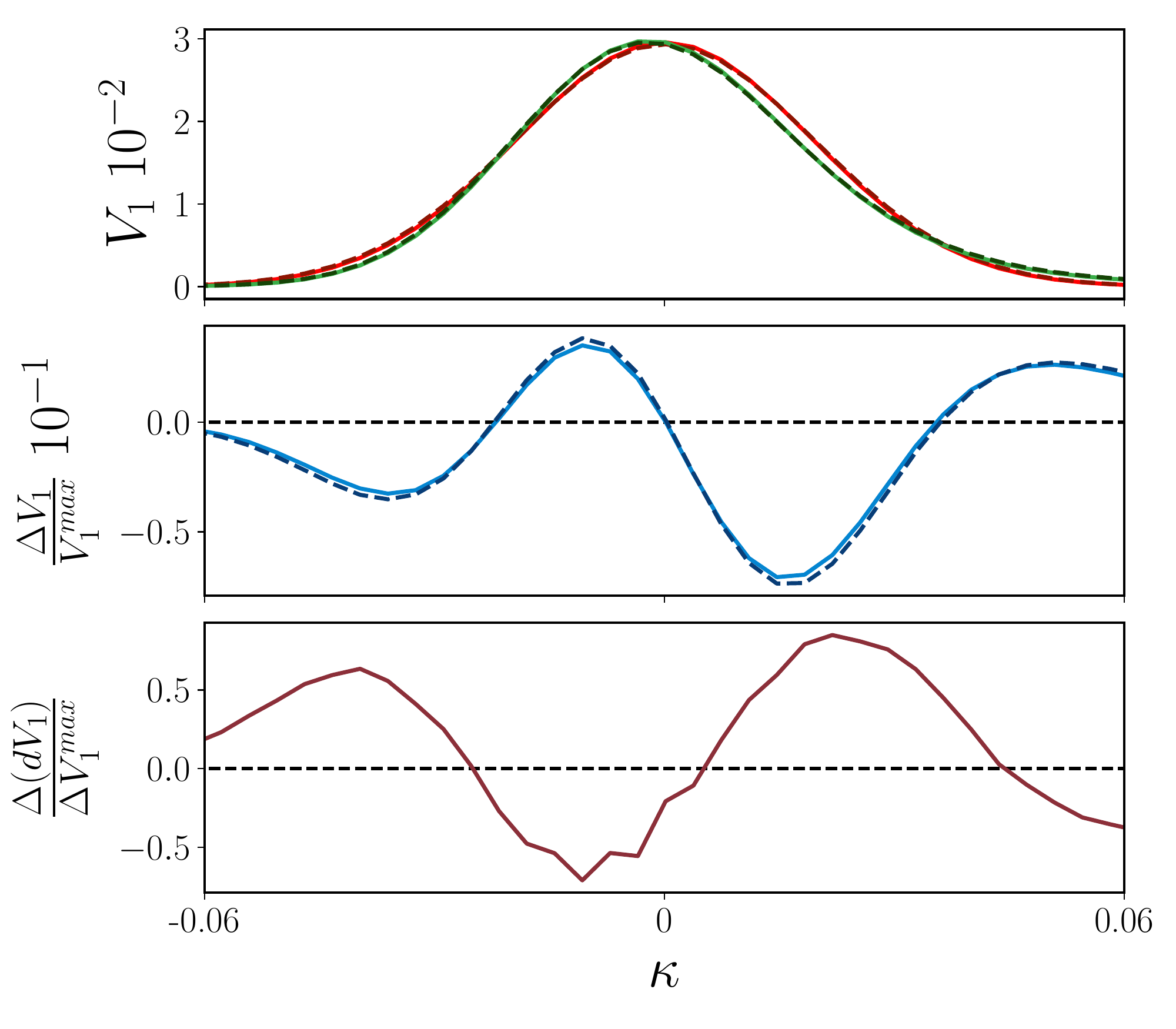}
	\end{subfigure}	
		\begin{subfigure}{0.49\linewidth}
		\includegraphics[width=\linewidth]{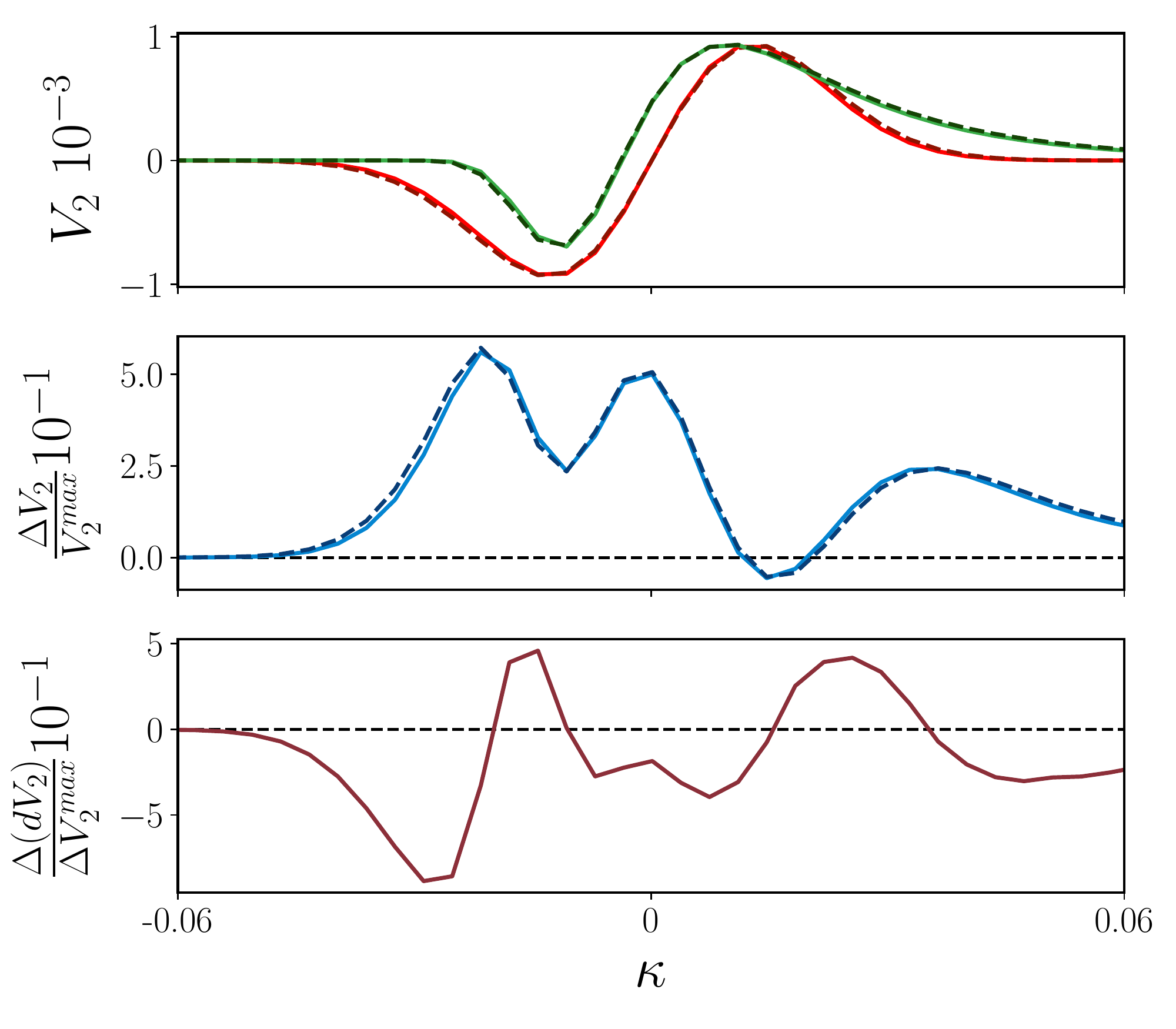}
	\end{subfigure}	
	\begin{subfigure}{0.49\linewidth}\includegraphics[width=\linewidth]{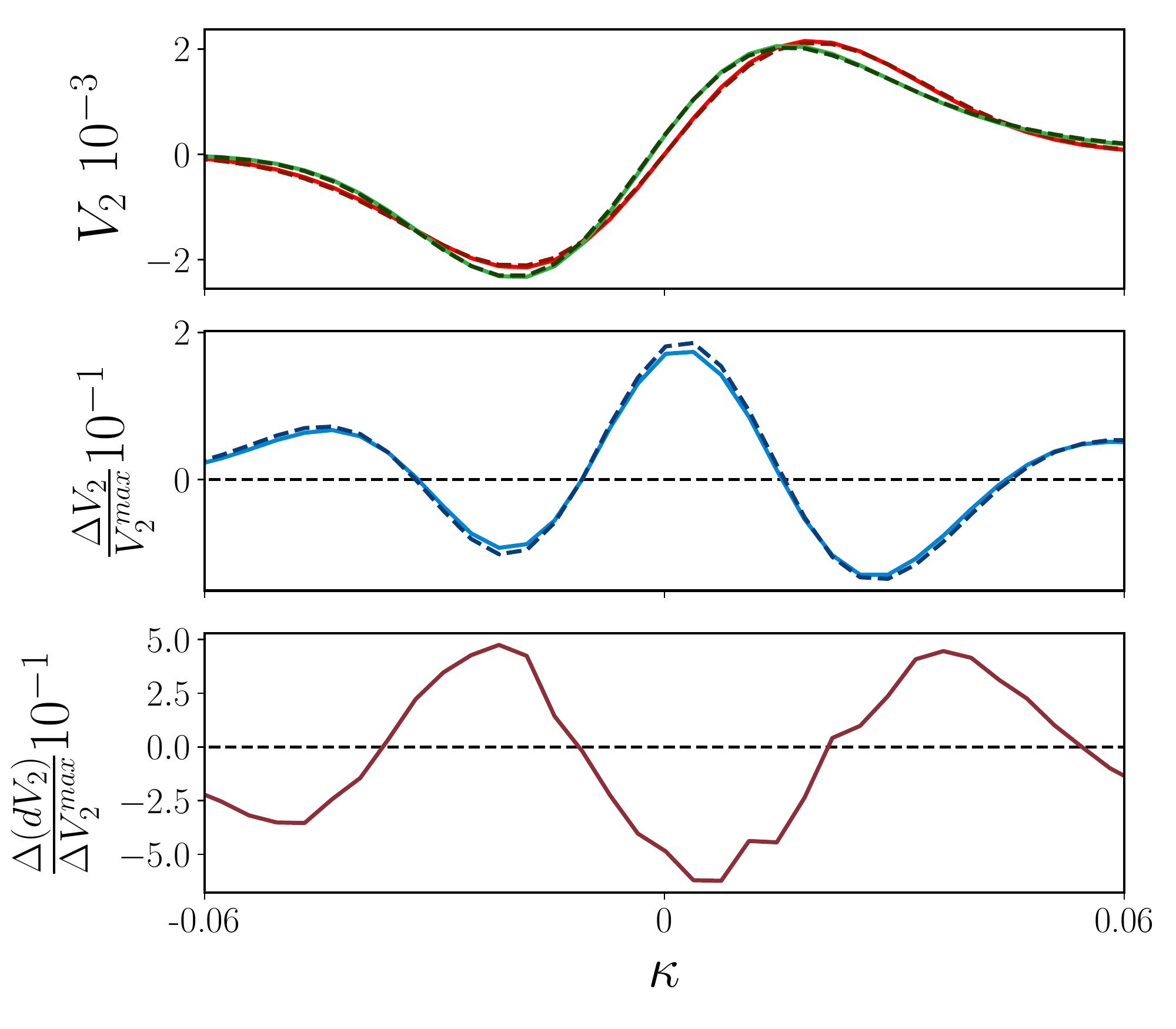}	 
	\end{subfigure}	
\caption{Noiseless ({\bf left column}) and noisy ({\bf right column}) MFs ({\bf top to bottom}: $V_0, V_1, V_2$) of the massive (solid) and massless (dashed) fiducial models, in comparison to that of the theoretical prediction for a GRF. Source galaxies are assumed to be at $z_{s}=1.0$ and all maps are smoothed with a one arcmin Gaussian window. In middle panel of each subplot, we show the fractional difference between the simulated and Gaussian MFs, demonstrating the non-Gaussian feature in MFs. In the bottom panel of each subplot, we show the difference in their parameter sensitivity,  ($d V^{\rm sim}- dV^{\rm GRF})/dV_{\rm max}^{\rm sim}$, where $d V=V^{\rm massive}-V^{\rm massless}$.}
\label{fig:GRF}
\end{figure}

We investigate the information content in MFs beyond Gaussian predictions. For a GRF, we can write down analytical formula for the MFs~\cite{hikage2006primordial,matsubara2010analytic}. For a direct comparison, we use the measured average moments of the power spectrum from our simulations as an input to the theoretical Gaussian prediction. We show in Fig.~\ref{fig:GRF} the MFs measured from our simulations to the theoretical Gaussian predictions. The latter are expected to add no additional information to two-point statistics. We use only $z_s=1$ maps for simplicity.

We show the fractional differences between the simulated and Gaussian MFs, for both fiducial models (middle panels).
We see deviations for all three MFs, demonstrating that MFs are indeed non-Gaussian due to nonlinear structure growth. In addition, we also investigate their sensitivity to the neutrino mass, by comparing $dV=V^{\rm massive}-V^{\rm massless}$ for simulated and Gaussian MFs (bottom panels) and find they differ in almost the full $\kappa$ range. These tests show that not only are the MFs non-Gaussian, but also their cosmological sensitivity are different from the Gaussian prediction. Therefore, we expect MFs to contain additional information beyond second-order statistics.

\subsection{Improvement from tomography}
 
\begin{figure}[!h]
	\centering
	\begin{subfigure}{0.49\linewidth}
		\includegraphics[width=\linewidth]{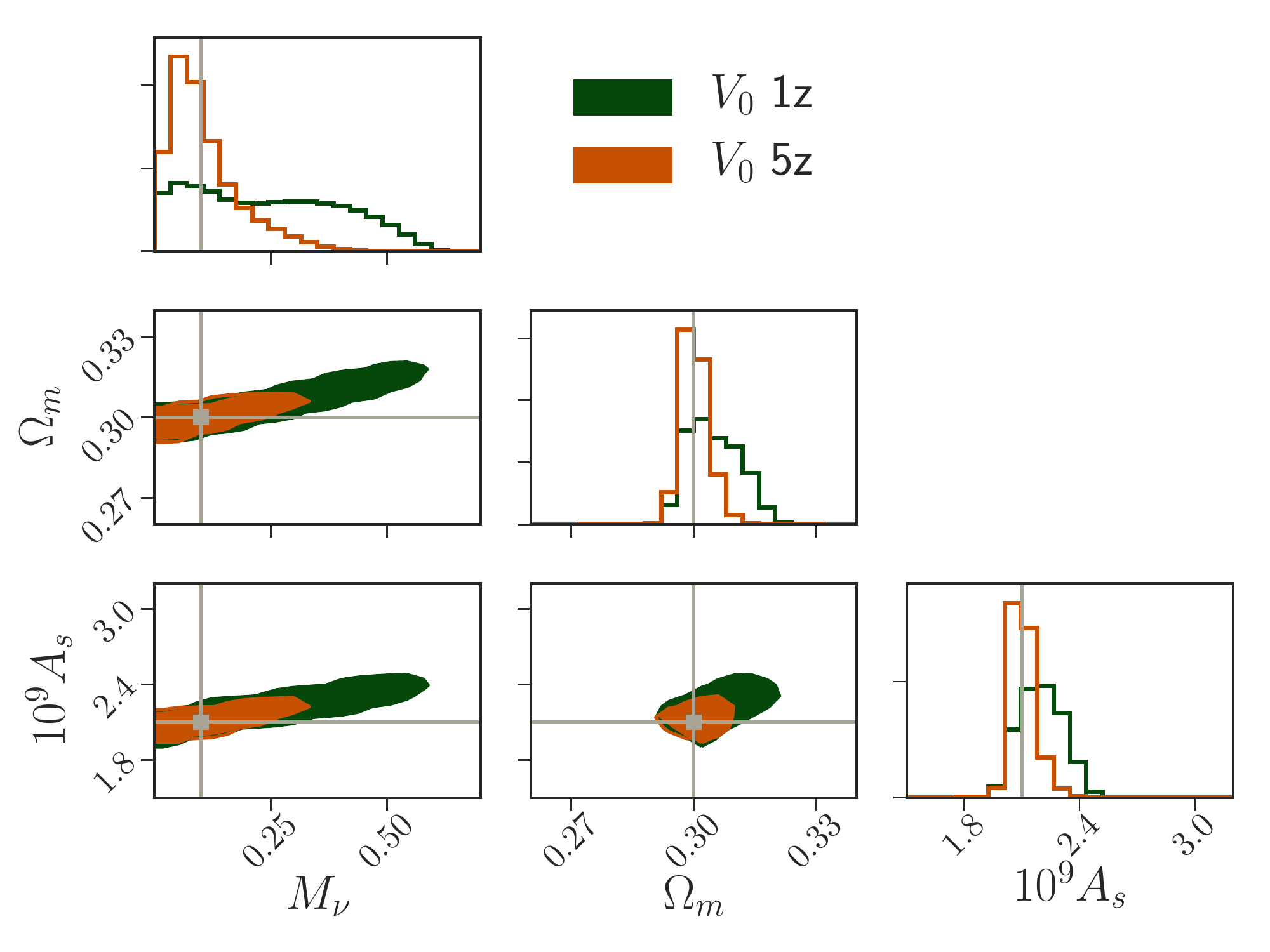}
	\end{subfigure}
	\begin{subfigure}{0.49\linewidth}
		\includegraphics[width=\linewidth]{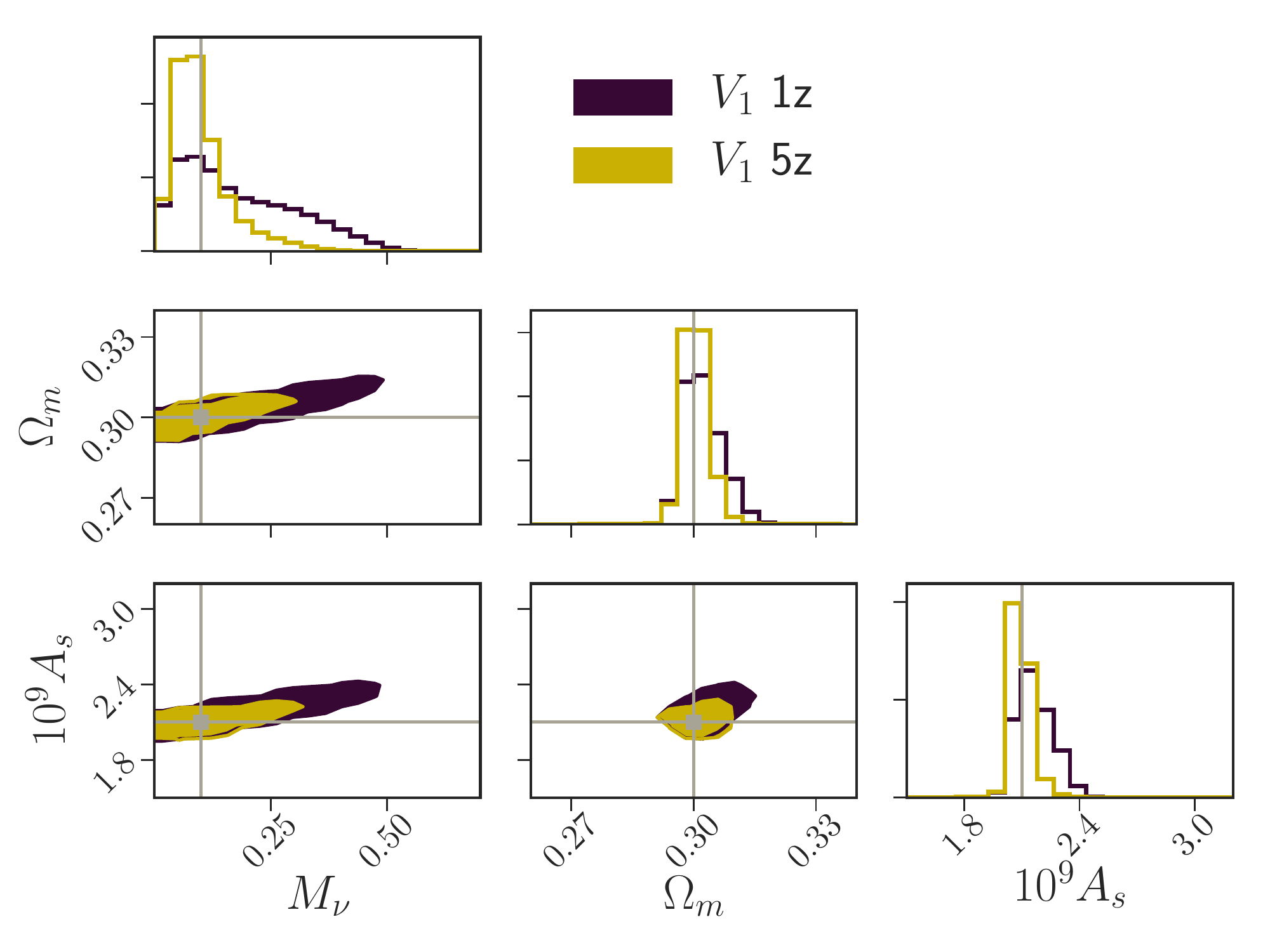}

	\end{subfigure}
	\begin{subfigure}{0.49\linewidth}
	\includegraphics[width=\linewidth]{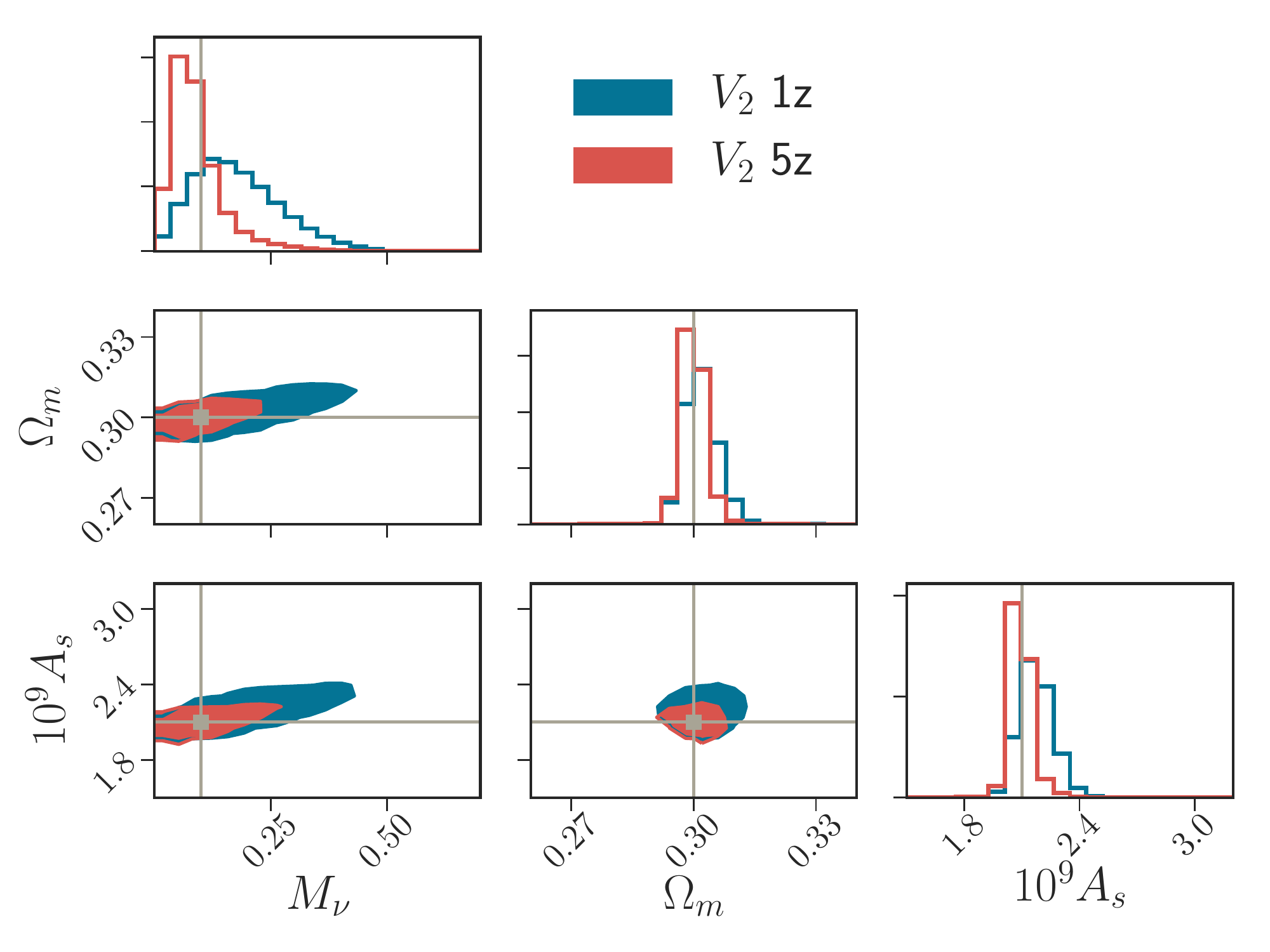}

	\end{subfigure}
	\begin{subfigure}{0.49\linewidth}
	\includegraphics[width=\linewidth]{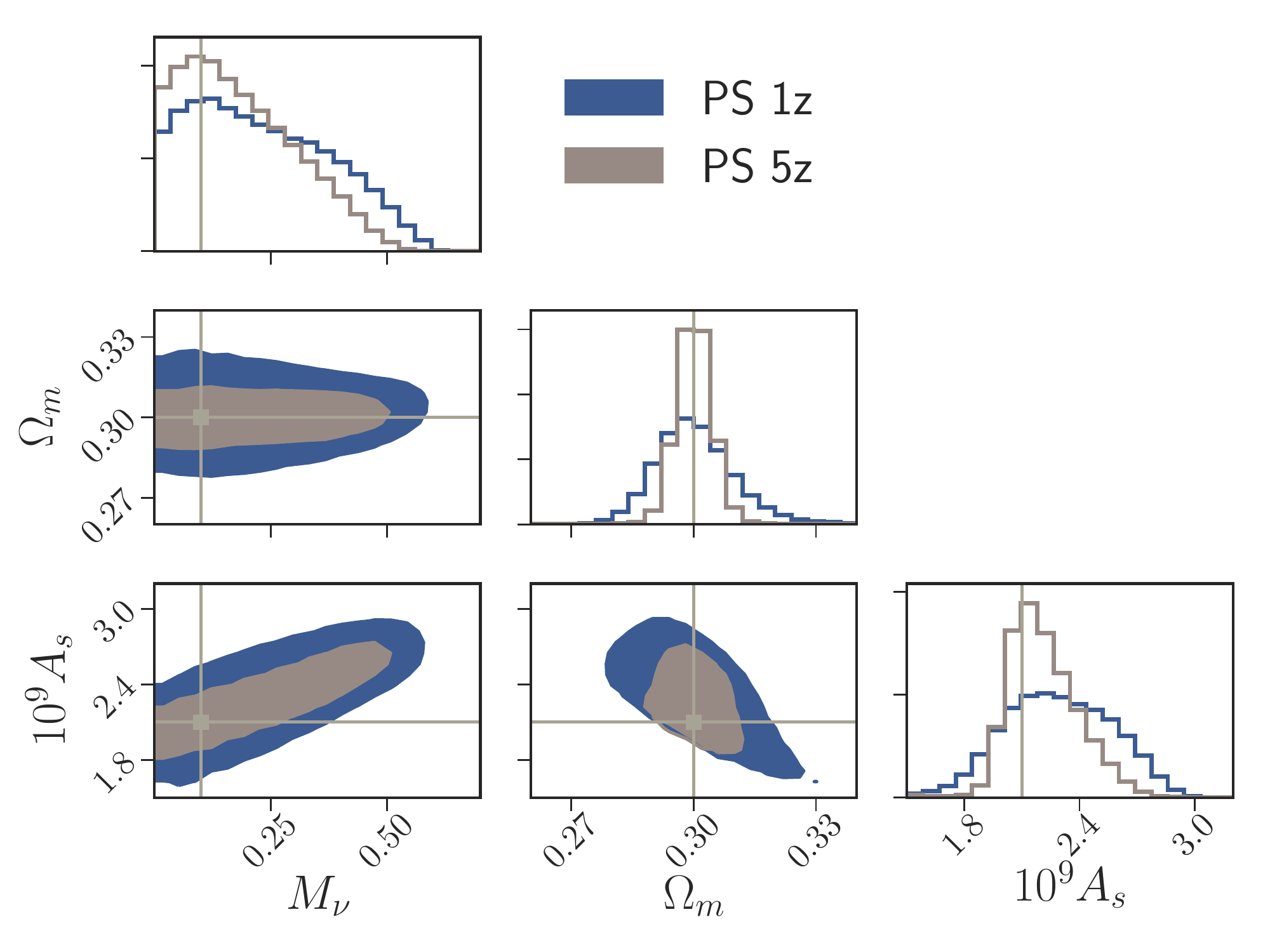}
	\end{subfigure}

\caption{95\% Constraints for the three MFs and the power spectrum (``PS'') from five redshift tomography (``5$z$'') and a single source redshift $z_s$=1.0  (``1$z$''), both with the same total galaxy number density. While tomography benefits the power spectrum in constraining $\Omega_m$, it is particularly powerful in tighten the $M_\nu$ errors for all three MFs, by at least a factor of two.}
	\label{fig:MFs_1z_5z}
\end{figure}

Redshift tomography, or splitting galaxy samples into multiple redshift bins, was proposed as a useful tool to capture the growth history ~\cite{giannantonio2016cmb,cai2012combining,li2018constraining,liu2018constraining, hu1999power,hannestad2006measuring,petri2016cosmology,hu2002dark,mancini2018testing}.  We study the additional constraining power from redshift tomography for both the power spectrum and MFs by comparing constraints from five redshift tomographic bins (the configuration adopted in our final analysis) to that from a single redshift $z_s=1$ with the same total galaxy number density, denoted as ``5$z$'' and ``1$z$'', respectively. 

In Fig.~\ref{fig:MFs_1z_5z} we show the 95$\%$ confidence contours for ``1$z$'' and ``5$z$'' for the three MFs and the power spectrum (``PS''). We find that tomography helps shrink the contours for all four statistics.  
We list in Table~\ref{table: 5z} the improvements quantified as the ratio of ``1$z$'' to ``5$z$'' marginalized 95\% errors. We find that while for the power spectrum tomography is particularly helpful in tightening the constraint on $\Omega_m$, MF constraints improve most significantly on $M_\nu$ by at least a factor of two.

\begin{table}[h!]
\centering
\begin{tabular}{cccc}
\multicolumn{1}{l}{}        & \multicolumn{1}{l}{}           & \multicolumn{1}{l}{$\sigma_{p}^{1z}/\sigma_{p}^{5z}$} & \multicolumn{1}{l}{}         \\ \hline
\multicolumn{1}{|c|}{$p$}   & \multicolumn{1}{c|}{$M_{\nu}$} & \multicolumn{1}{c|}{$\Omega_{m}$}                                           & \multicolumn{1}{c|}{$A_{s}$} \\ \hline
\multicolumn{1}{|c|}{$V_0$} & \multicolumn{1}{c|}{2.55}        & \multicolumn{1}{c|}{2.07}                                                     & \multicolumn{1}{c|}{2.09}      \\ \hline
\multicolumn{1}{|c|}{$V_1$} & \multicolumn{1}{c|}{2.50}        & \multicolumn{1}{c|}{1.63}                                                     & \multicolumn{1}{c|}{2.21}      \\ \hline
\multicolumn{1}{|c|}{$V_2$} & \multicolumn{1}{c|}{2.02}        & \multicolumn{1}{c|}{1.46}                                                     & \multicolumn{1}{c|}{1.79}      \\ \hline
\multicolumn{1}{|c|}{PS}    & \multicolumn{1}{c|}{1.23}        & \multicolumn{1}{c|}{2.16}                                                     & \multicolumn{1}{c|}{1.65}      \\ \hline
\end{tabular}
\caption{Improvement on the three cosmological parameters from tomography, for the three MFs and the power spectrum (``PS''), represented as ratios of the marginalized $95\%$ errors from a single redshift bin (``1$z$'') to that from five tomographic redshift bins (``5$z$'').}
\label{table: 5z}
\end{table}

\subsection{Joint constraints}
\begin{figure}[!h]
	\centering
	\includegraphics[width=\linewidth]{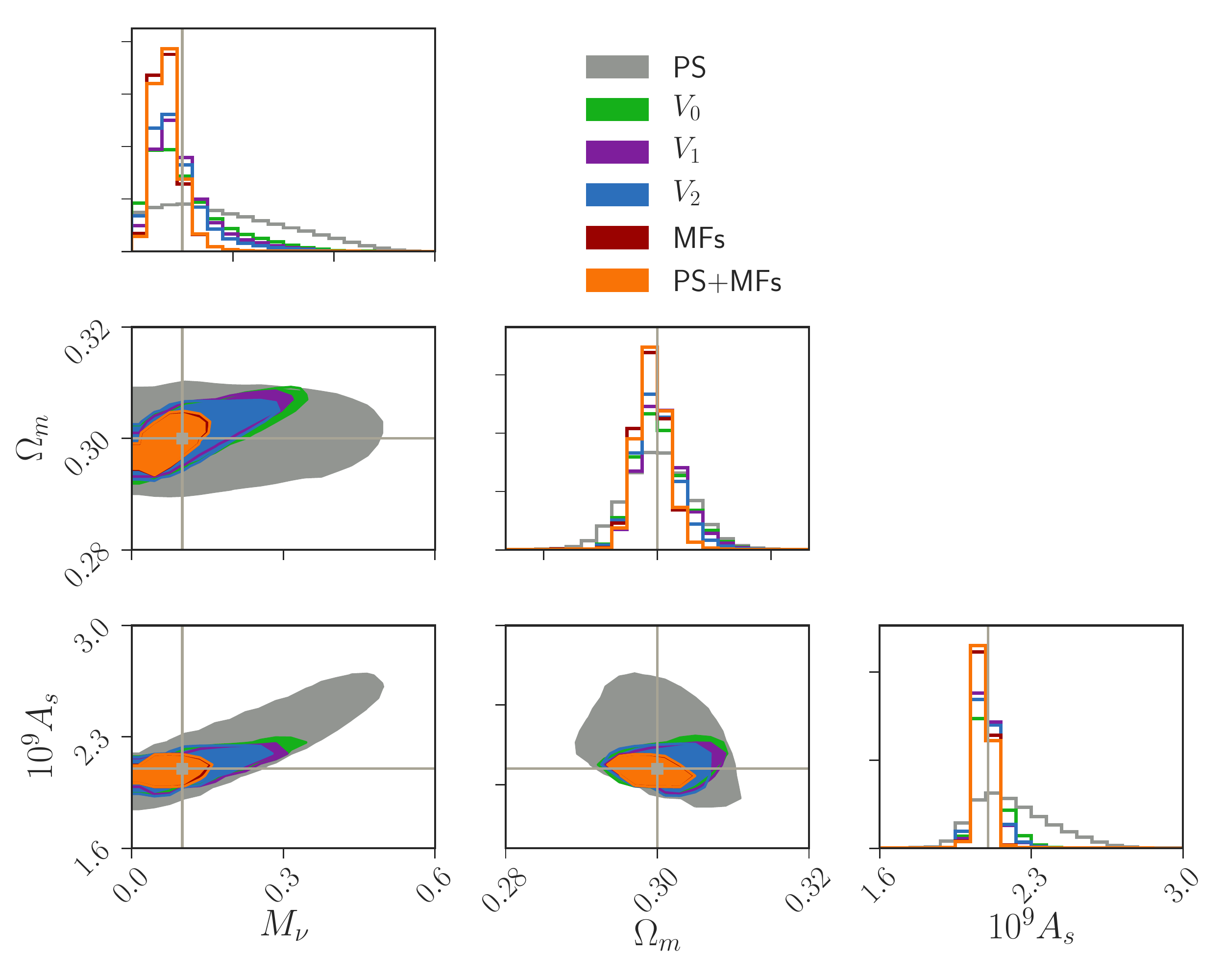}
\caption{$95\%$ confidence contours from the power spectrum (``PS'') (grey), three Minkowski Functionals $V_{0}$ (green), $V_{1}$ (purple) and $V_{2}$ (blue), three MFs combined (red), and  PS and MFs combined (orange). We assume an LSST-like survey with five redshift tomography $z_s = 0.5, 1.0, 1.5, 2.0, 2.5$, with full covariance included. The fiducial values are shown in grey lines.}
\label{fig:5z_constraints}
\end{figure}
Finally, we inspect the constraints on [$M_{\nu}$, $\Omega_{m}$,  $A_{s}$] from $V_{0}$, $V_{1}$, $V_{2}$, and the power spectrum, as well as from combining all three MFs and combining MFs and the power spectrum. We assume five redshift tomography in all cases. The 95\% confidence contours from these descriptors are shown in Fig.~\ref{fig:5z_constraints}. 

In all 2D projected likelihood planes, [$\Omega_{m}$, $M_{\nu}$], [$A_{s}$, $M_{\nu}$] and [$A_{s}$, $\Omega_{m}$], each MF contour alone already appears tighter than that of the power spectrum. All the MF contours show different degeneracy from that of the power spectrum, a result of the high-order information they contain. Combining all three MFs further improves the constraints from individual MF, especially on  neutrino mass. Adding the power spectrum brings negligible improvement to that of the MFs, showing that MFs already contain most of the two-point information.  We quantify in Table~\ref{5z_all} the improvement from each descriptor using a ratio of the marginalized $95\%$ errors from the power spectrum to that of each descriptor in interest.

\begin{table}
\centering
\begin{tabular}{cccc}
\multicolumn{1}{l}{}          & \multicolumn{1}{r}{}           & \multicolumn{1}{l}{$\sigma_{p}^{PS}/\sigma_{p}$} & \multicolumn{1}{l}{}         \\ \hline
\multicolumn{1}{|c|}{$p$}     & \multicolumn{1}{c|}{$M_{\nu}$} & \multicolumn{1}{c|}{$\Omega_{m}$}                  & \multicolumn{1}{c|}{$A_{s}$} \\ \hline
\multicolumn{1}{|c|}{$V_{0}$} & \multicolumn{1}{c|}{1.67}      & \multicolumn{1}{c|}{1.30}                          & \multicolumn{1}{c|}{2.83}    \\ \hline
\multicolumn{1}{|c|}{$V_{1}$} & \multicolumn{1}{c|}{2.46}      & \multicolumn{1}{c|}{1.49}                          & \multicolumn{1}{c|}{3.95}    \\ \hline
\multicolumn{1}{|c|}{$V_{2}$} & \multicolumn{1}{c|}{2.79}      & \multicolumn{1}{c|}{1.58}                          & \multicolumn{1}{c|}{3.57}    \\ \hline
\multicolumn{1}{|c|}{MFs}     & \multicolumn{1}{c|}{4.82}      & \multicolumn{1}{c|}{2.01}                          & \multicolumn{1}{c|}{5.59}    \\ \hline
\multicolumn{1}{|c|}{MFs+PS}  & \multicolumn{1}{c|}{4.87}      & \multicolumn{1}{c|}{2.06}                          & \multicolumn{1}{c|}{5.76}    \\ \hline
\end{tabular}
\caption{Ratios of marginalized $95\%$ errors from the power spectrum $\sigma^{PS}$ to that from each descriptor $\sigma^{p}$ in interest. We assume galaxy noise and redshift distribution for an LSST-like survey, with five tomographic redshift bins $z_s = 0.5, 1.0, 1.5, 2.0, 2.5$.}
\label{5z_all}
\end{table} 

The marginalized constraint on neutrino mass from three MFs combined improves that from the power spectrum alone by a factor of 4.82. Comparing the constraining power from the three MFs, we find that $V_{2} \gtrsim V_{1} > V_{0}$, similar to findings in previous work~\cite{ducout2012non,kratochvil2012probing, novaes2016local}. This is because both perimeter ($V_1$) and genus ($V_2$) are sensitive to not only the overall distribution of under/over-density, as probed by area ($V_0$), but also their spatial information. 

Our results show that MFs will no doubt be a powerful tool for future lensing surveys to constrain neutrino mass.

\section{Summary}
\label{sec:conclusions}
We study the effect of massive neutrinos on weak lensing Minkowski Functionals, which can capture non-Gaussian information beyond  two-point statistics. We forecast the improved constraints from MFs on cosmological parameters, including the neutrino mass sum ($M_{\nu}$), matter density ($\Omega_{m}$) and primordial power spectrum amplitude ($A_{s}$), compared to those expected from the weak lensing power spectrum. We use the \texttt{MassiveNuS} simulations with five tomographic source redshifts and assume  galaxy distribution and noise properties of an LSST-like survey. We also explore the importance of redshift tomography. We show constraints from three individual MF, them combined, and MFs and the power spectrum combined. Our main conclusions are:

\begin{itemize}
\item The three MFs are sensitive to massive neutrinos. Neutrino signatures are redshift dependent (Fig.~\ref{fig:MFs_signature}) and are different from the Gaussian prediction (Fig.~\ref{fig:GRF}). Therefore, MFs contain non-Gaussian information beyond the power spectrum.

\item Redshift tomography, or multiple redshift bins, is a powerful technique to constrain neutrino mass, as neutrinos have distinct evolutionary signature than other parameters. Compared to using a single redshift bin, using tomography we find more than a factor of two improvements on $M_{\nu}$ (95\% C.L.) for all three MFs. While the improvement on neutrino mass from tomography is milder for the power spectrum~($\approx$20\%), that on $\Omega_m$ is much stronger, by factor of two (Fig.~\ref{fig:MFs_1z_5z}). 

\item Each MF alone already outperforms the power spectrum in constraining cosmology $[M_{\nu}$, $\Omega_{m}$, $A_{s}]$. Combining the three MFs further improves the constraint by more than a factor of four, in comparison to that from the power spectrum alone (Fig.~\ref{fig:5z_constraints}). Joint analysis of the MFs and the power spectrum returns similar constraints as the MFs. The relative constraining power of the MFs are: $V_{2} \gtrsim V_{1} > V_{0}$.

\end{itemize}

In summary, Minkowski Functionals allow us to probe additional information that is missing in traditional two-point analyses. They are a powerful tool to constrain the neutrino mass sum for upcoming cosmological surveys. In our work, however, we only considered errors from galaxy shape measurements. To realize the full potential of MFs, follow-up work should address the impact of other systematics, for example, baryonic physics, photometric redshift biases, multiplicative bias in galaxy shapes and intrinsic alignments of galaxies. While it is unrealistic to assume MFs to be immune to these systematics, we expect these systematics to impact MFs differently than the power spectrum, and joining the two statistics will also bring the possibility to jointly model cosmology, astrophysics, and other nuisance parameters. 
Other issues to be consider include the degeneracy of neutrino mass with  others cosmological parameters, such as interacting dark matter and time-dependent dark energy. These are beyond the scope of our work, but are necessary steps to take before we can fully trust the analysis results. Finally, we expect the most powerful constraints to come from combining weak lensing with other cosmological probes including CMB temperature and polarization, CMB lensing, Baryon Acoustic Oscillation, and Lyman alpha forest.

\acknowledgments 
We thank Will Coulton, Zack Li, Francisco Villaescusa-Navarro, Ben Wandelt for useful discussions. 
GAM is supported by the CAPES Foundation of the Ministry of Education of Brazil fellowships. This work is supported by an NSF Astronomy and Astrophysics Postdoctoral Fellowship (to JL) under award AST-1602663. ZH and JMZM were supported by NASA grant 80NSSC18K1093. CPN is supported by the FAPERJ Brazilian funding agency. We acknowledge support from the WFIRST project. We thank New Mexico State University (USA) and Instituto de Astrofisica de Andalucia CSIC (Spain) for hosting the Skies \& Universes site for cosmological simulation products.
This work used the Extreme Science and Engineering Discovery Environment (XSEDE), which is supported by NSF grant ACI-1053575. GAM thanks the cordial host at Department of Astrophysical Sciences at Princeton University 
and Florida Space Institute at University of Central Florida, where part of this project was developed.

\bibliographystyle{JHEP}
\bibliography{main}

\end{document}